\documentclass[aps,prl,preprint]{revtex4}
\usepackage{graphicx}
\usepackage{float}
\usepackage{amsfonts}
\usepackage{amsmath}
\usepackage{amssymb}
\usepackage{enumerate}
\usepackage{color}
\usepackage{mciteplus}
\usepackage{natbib}

\usepackage{epstopdf} 

\newcommand{\bra}[1]{\left\langle{#1}\right\vert}
\newcommand{\ket}[1]{\left\vert{#1}\right\rangle}
\newcommand{\be}{\begin{equation}}
\newcommand{\ee}{\end{equation}}

\begin{document}

\title{Quantum Statistics of Surface Plasmon Polaritons \\ in Metallic Stripe Waveguides}





\author{Giuliana Di Martino\,$^a$,$^1$ Yannick Sonnefraud\,$^a$,\footnote{Email: y.sonnefraud@imperial.ac.uk}$^{,1}$ St\'ephane K\'ena-Cohen\,$^a$,$^1$ Mark Tame\,$^a$,$^2$ \c{S}ahin K. \"Ozdemir,$^3$ M. S. Kim,$^{2}$ Stefan A. Maier\footnote{Email: s.maier@imperial.ac.uk}$^{,1}$}

\affiliation{$^1$Imperial College London, Blackett Laboratory, Experimental Solid State Group, SW7 2AZ London UK \\
$^2$Imperial College London, Blackett Laboratory, Quantum Optics and Laser Science Group, SW7 2AZ London UK \\
$^3$Department of Electrical and Systems Engineering, Washington University, St. Louis, MO 63130, USA \\
$^a$\,{\rm {\footnotesize These authors contributed equally.}}$\qquad$}


\begin{abstract}
Single surface plasmon polaritons are excited using photons generated \emph{via} spontaneous parametric down-conversion. The mean excitation rates, intensity correlations and Fock state populations are studied. The observed dependence of the second order coherence in our experiment is consistent with a linear uncorrelated Markovian environment in the quantum regime. Our results provide important information about the effect of loss for assessing the potential of plasmonic waveguides for future nanophotonic circuitry in the quantum regime.  
\end{abstract}


\maketitle

Surface plasmon polaritons (SPPs) are highly confined electromagnetic excitations coupled to electron charge density waves propagating along a metal-dielectric interface. A significant effort is currently being devoted to the study of their unique light-matter properties and to their use in optoelectronic devices exhibiting sub-wavelength field confinement~\cite{Takahara1997, Ebbesen2008}. Most recently there has been a growing excitement among researchers about the prospects for building plasmonic devices that operate faithfully at the quantum level. Indeed, the hybrid nature of SPPs and the potential for strong coupling to emitter systems \emph{via} intense, highly confined fields~\cite{Chang2006,Akimov2007} offers new opportunities for the quantum control of light~\cite{Zuloaga2009,Artuso2011}. The main hindrance to the use of SPPs in practical devices is, however, their lossy character. Still, recent work has shown that SPPs can maintain certain quantum properties of their exciting photon field, with the demonstration of assisted transmission of entangled photons~\cite{Altewischer2002,Moreno2004}, energy-time entanglement~\cite{Fasel2005}, quantum superposition~\cite{Fasel2006}, quadrature squeezing~\cite{Huck2009}, wave-particle duality~\cite{Kolesov2009} and single plasmon detection~\cite{Heeres2010}. These results suggest that many principles of quantum optics can be transferred to the field of plasmonics, enabling novel devices such as single-photon switches to be realized~\cite{Chang2007}. Despite recent progress in using quantum optical techniques to study plasmonic systems, adapting them to realistic structures will require a much more detailed understanding of the quantum properties of SPPs when loss is present. This is the central focus of our investigation and represents an area so far lacking an in-depth experimental study. Understanding how loss affects the quantum properties of SPPs may open up a route toward the realistic design and fabrication of nanophotonic plasmonic circuits for quantum information processing.

In this letter, we characterize the effects of loss on the quantum statistics of heralded waveguided SPPs. Using single photons produced by type-I spontaneous parametric down conversion (SPDC)~\cite{Hong1986,Burnham1970}, we excite quanta of leaky SPPs in thin metallic stripe waveguides, one of the fundamental building blocks for plasmonic circuits.~\cite{Berini2001,Lamprecht2001,Tame2008,Zia2005} We measure the second-order quantum coherence function, $g^{(2)}(\tau)$, Fock state populations and mean excitation count rates for a range of different waveguide lengths. We find that the mean excitation rate follows the classical intensity rate as the waveguide length increases, but that the second-order quantum coherence remains markedly different from that expected in the classical regime and keeps a constant value. The measured dependence in our experiment is consistent with a linear uncorrelated Markovian environment in the quantum regime~\cite{Tame2008}. Our results provide important information about the effect of loss for assessing the realistic potential of building plasmonic waveguides for nanophotonic circuitry that operates faithfully in the quantum regime.

\begin{figure}[t]
\includegraphics[width=14cm]{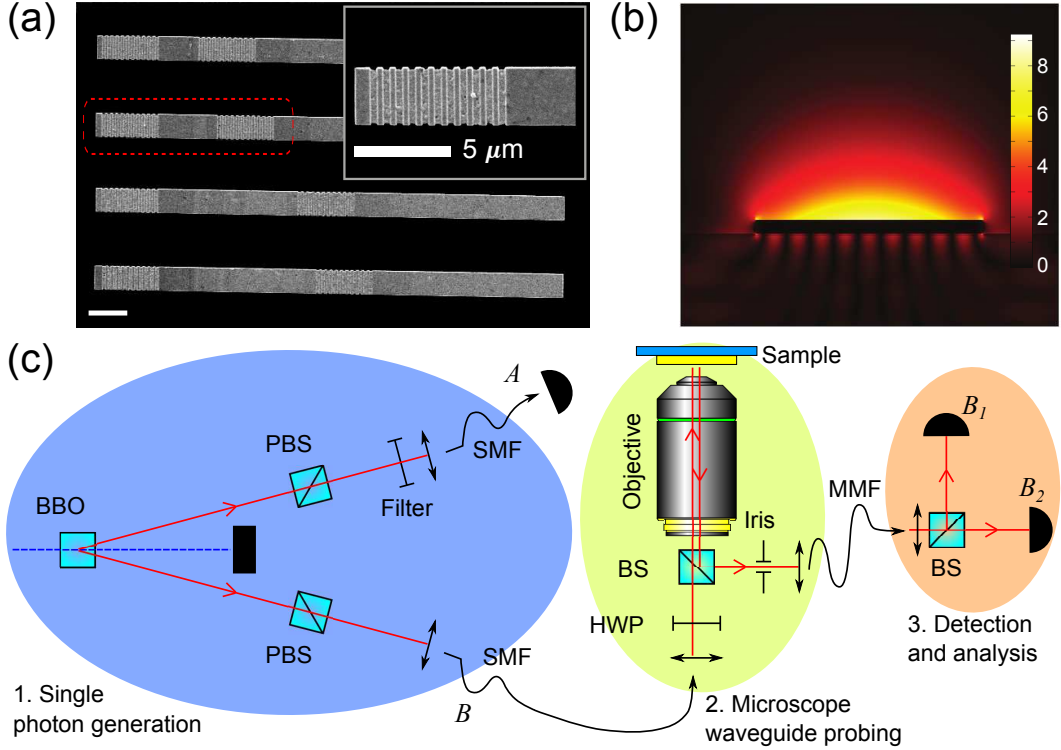}
\caption{Experimental configuration. (a)~Scanning electron microscope image of a selection of waveguide lengths from 5~$\mu$m to 20~$\mu$m. All waveguides have been explored in the quantum regime, however a further time-domain analysis of the quantum statistics has been performed for the 7.5~$\mu$m waveguide highlighted by the dashed red box. Inset: detail of one of the in/out-coupling gratings. The scale bars denote 5~$\mu$m. (b)~Fundamental SPP mode in our stripe waveguide -- electric field profile along the cross section of the waveguide, calculated using the Finite Element Method (FEM) for an infinitely long waveguide. (c)~Schematic of the experimental setup including a single-photon source stage, waveguide probing stage and final analysis stage.}
\label{figure1}
\end{figure}

	
The structures studied consist of 3~$\mu$m wide, 150~nm thick gold stripes fabricated by electron-beam lithography on glass coated with 23~nm of indium tin oxide. Input and output gratings with a periodicity of 680~nm were then etched into the waveguides using focused ion-beam milling. To investigate the effects of loss on SPP excitations, the separation between gratings was varied between 5~$\mu$m and 30~$\mu$m in steps of 2.5~$\mu$m. A scanning electron microscope image of a selection of our waveguides is shown in Fig.~\ref{figure1}(a). These asymmetric waveguides support a number of leaky~\cite{Zia2005} and bound~\cite{Berini2001} quasi-TM guided plasmon modes. Here, the grating periodicity (see inset of Fig.~\ref{figure1}(a)) was chosen to couple effectively to the leaky-modes supported by the structure (those with highest field intensity at the gold-air interface). Two such modes exist in the waveguides and the lowest order mode, shown in Fig.~\ref{figure1}(b), possesses by far the lowest losses of the complete set of supported modes. Indeed, the leaky-modes of asymmetric waveguides are most often used as plasmonic circuitry components due to their low losses as compared to the bound modes.~\cite{Ebbesen2008} 

\begin{figure}[t]
\includegraphics[width=16cm]{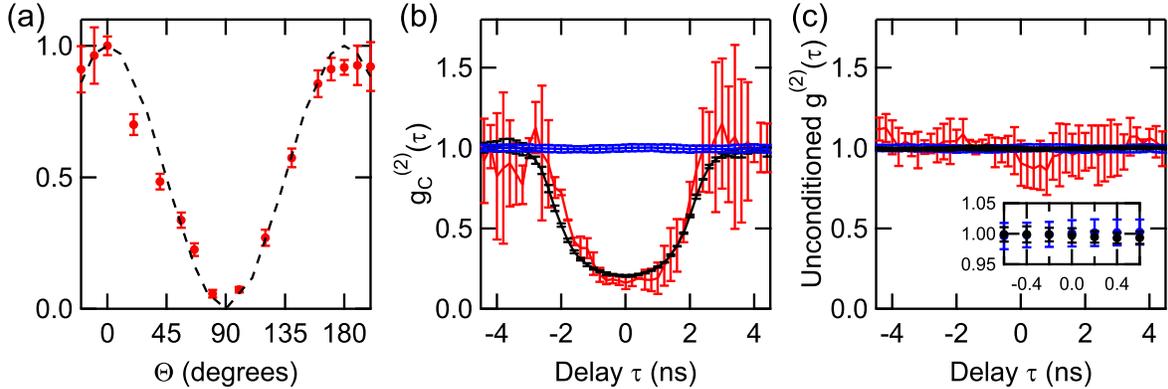}
\caption{Intensity dependence and second-order quantum coherence. (a) Normalized coincidence rate coupled out from a 7.5$\mu$m waveguide, dependent on the polarization angle $\Theta$ of the beam incident on the in-coupling grating (red) and theoretically expected ${\text{cos}}^2(\Theta)$ dependence (black dashed line). (b) Conditional second-order quantum coherence function, $g^{(2)}_c(\tau)$, for the down-converted light in mode $B$ before the waveguide (black), along with the out-coupled light when single photons are injected into a waveguide of length 7.5$\mu$m (red). The classical limit is illustrated by the blue data points, corresponding to the unconditioned second-order quantum coherence, $g^{(2)}(\tau)$, for an attenuated laser injected in the waveguide. (c) $g^{(2)}(\tau)$ for the down-converted light in mode $B$ (black) and injected into the waveguide (red). Blue: $g^{(2)}(\tau)$ for the attenuated laser injected in the waveguide. The inset shows a magnified region of the graph, omitting the red curve.}
\label{figure2}
\end{figure}

Fig.~\ref{figure1}(c) shows a schematic of the experimental setup used to conduct our investigation. To create down-converted photon pairs at 808~nm, a 100~mW CW, $\lambda$=404~nm laser is focused onto a BBO crystal cut for type-I SPDC. Phase matching conditions~\cite{Hong1986,Burnham1970} lead to the photons from a given pair being emitted into antipodal points of a forward-directed cone with an opening angle of 6$^\circ$ (Fig.~\ref{figure1}(c) shows that the antipodal points chosen are in-plane). Polarizing beam splitters (PBSs) are placed in the path of the down-converted beams to remove any parasitic light with the incorrect polarization. Each beam is then injected into a single mode fiber (SMF). A filter is placed on path $A$ to spectrally select out the down-converted photons (see Supplementary Information). One of the fibers is directly connected to a silicon avalanche photodiode detector (APD $A$ in Fig.~\ref{figure1}(c)) which monitors the arrival of one photon from a given SPDC pair. The second fiber is used to channel the other photon from the pair to the sample in the microscope. In our experiment, APD $A$ acts as a trigger device, where the detection of a photon in fiber $A$ `heralds'  the presence of a photon in the second fiber $B$ (see Supplementary Information). This system allows for the generation of heralded photons at a rate of 10$^6$s$^{-1}$. After collimation, the polarization of the generated single photon is adjusted using a half waveplate (HWP) and then focused onto the in-coupling grating of the waveguide probed. At the grating, the generated photons are converted into SPPs~\cite{Tame2008} due to phase-matching conditions (see Supplementary Information). These propagate along the waveguide until they reach the out-coupling grating, at which point they are converted back into light. The output is selected by an iris and injected into a multimode fiber (MMF) for analysis. The dependence of the output light on the input polarization (as shown in Fig.~\ref{figure2}(a)) confirms that the collected light originates exclusively from out-coupled SPPs. Finally, the multimode fiber directs the output to a Hanbury Brown and Twiss interferometer (BS and detectors $B_1$ and $B_2$) used to measure the second-order quantum coherence function, $g^{(2)}(\tau)$.~\cite{Brown1956} All of the photon detection events are time-tagged (Hydraharp 400, PicoQuant GmbH). This allows measurements involving detectors $B_1$ and $B_2$ to be conditioned on the detection of a photon at detector $A$, with appropriate delays, ensuring that only correlations arising from the injection of single photons into the waveguide are measured.



We first measure the second-order quantum coherence function, $g^{(2)}(\tau)$, as a function of time delay $\tau$ between detectors
$B_1$ and $B_2$. This coherence function is a measure of the correlation of the intensity of a field at a time $t=0$ and at a later time $t=\tau$ for a fixed position. By measuring $g^{(2)}(0)$ for a given field, we can determine whether or not it is in the nonclassical regime ($g^{(2)}(0)<1$). In particular, for number states $\ket{n}$, if $g^{(2)}(0)<1/2$ is measured in an experiment, we can be confident that the field is within the single excitation regime (see Supplementary Information).

A beamsplitter is used to symmetrically split the field in mode $B$ into modes $B_1$ and $B_2$. In this case one can show that the definition of $g^{(2)}(\tau)$ (cf. Equation~1 in Supplementary Information) is equivalent to~\cite{Loudon2000,Thorn2004} 
\begin{equation}
g^{(2)}(\tau)=\frac{\langle \hat{E}_{B_1}^-(0)\hat{E}_{B_2}^-(\tau)\hat{E}_{B_2}^+(\tau)\hat{E}_{B_1}^+(0)\rangle}{\langle \hat{E}_{B_1}^-(0)\hat{E}_{B_1}^+(0)\rangle \langle \hat{E}_{B_2}^-(\tau)\hat{E}_{B_2}^+(\tau)\rangle}\equiv \frac{N_{B_1B_2}}{N_{B_1}N_{B_2}}\left( \frac{T}{\Delta t}\right),
\label{g22}
\end{equation}
where $\hat{E}^+(t)$ is the electric field operator, $T$ the averaging (integration) time of the measurement, $N_{B_1B_2}$ is the number of coincidence detections at detectors $B_1$ and $B_2$ within a coincidence time window $\Delta t$, and $N_{B_1}$ and $N_{B_2}$ are the number of independent detections at detectors $B_1$ and $B_2$ respectively. All detections at $B_2$ are delayed by time $\tau$. In our experiment, we use both an attenuated laser source ($\lambda$=785~nm) with Eq.~\ref{g22} used to calculate the second-order quantum coherence function and the single-photon source, where all measurements are conditioned on the detection of a photon in mode $A$. Therefore in this second case we use the conditional form of Eq.~\ref{g22}, given by~\cite{Razavi2009}

\begin{equation}
g^{(2)}_c(\tau)=\frac{N_{A}N_{A B_1 B_2}}{N_{A B_1}N_{A B_2}},
\label{g23}
\end{equation}
where $N_{A B_1B_2}$ is the number of coincidence detections at detectors $A$, $B_1$ and $B_2$, with detections at $B_1$ and $B_2$ occurring within a coincidence time window $\Delta t$ centered on the detection at $A$. $N_{A B_1}$ is the number of coincidence detections at detectors $A$ and $B_1$ within the coincidence time window $\Delta t$ and similarly for  $N_{A B_2}$. $N_{A}$ is the number of independent detections at detector $A$. All measurements are taken over an integration time $T$, which does not appear explicitly in Eq.~\ref{g23}. The value of $g^{(2)}_c(\tau)$ at zero time delay provides us with a measure of conditioned single-arm statistics 
in mode $B$~\cite{Footnote1}. Additional information about the quality of the field intensity correlations can then be obtained by measuring $g^{(2)}_c(\tau)$ over a range of different time delays $\tau$~\cite{Razavi2009}. We have performed such a time-domain analysis in our experiment for the out-coupled light from single-photons injected into a fixed waveguide length of $7.5\mu$m. The results from the single-photon source before and after the waveguides are shown in Fig.~\ref{figure2}(b) (resp. black and red curves). The time window $\Delta t$ used for the measurement is 2~ns, and the integration times are adjusted to obtain reasonable error bars. The statistics obtained after the waveguide are identical to those of the source itself. One can see that the value of $g^{(2)}_c(0)$ is $<0.5$ in both cases which demonstrates that we are in the single plasmon excitation regime. Note that $g^{(2)}_c(0)$ is not identically zero in either case. The finite value, however, originates solely from accidental coincidences (see Supplementary Information and Ref.~[23]).
In Fig.~\ref{figure2}(c) we show the \textit{unconditioned} $g^{(2)}(\tau)$ as a function of time-delay for the single photon source only and after a 7.5$\mu$m length waveguide (resp. black and red). This plot shows the vital role of the detection of photons in mode A for the conditional measurements of the statistics of the out-coupled light from the waveguides. Without this `heralding' of the photons, the statistics of the light arriving at the detectors are those of a thermal field~\cite{Razavi2009}. It should be noted that the theoretically expected peak of $g^{(2)}(0)=2$ for a thermal field is challenging to observe in quantum optics experiments~\cite{Blauensteiner2009}, as instead of reaching the value of 2, its height above unity is effectively proportional to the ratio of the coherence time of the single-photon source to the response time of the detection, which in our experiment is $\sim 10^{-5}$.


We now turn our attention to the effect of losses in the single excitation regime. As the SPPs propagate along the waveguide, the finite conductivity of the metal results in ohmic losses, while radiation into the substrate and surface roughness result in radiative losses~\cite{Zayats2005}. For a reasonably smooth waveguide surface and a thick gold layer, ohmic losses are the main source of damping. Here we operate far enough from the plasmon wavelength for the free electron approximation to hold and we thus expect a linear loss model with uncorrelated Markovian noise to be valid. In this context we expect that for number states $\ket{n}$ the quantum observables that make up $g^{(2)}$ transform the numerator of Eq.~\ref{g22} as $n(n-1) \to \eta^2~n(n-1)$ and the denominator as $n \to \eta~n$, where $\eta$ is the total loss over the length of the waveguide~\cite{Tame2008}. Thus, for this particular loss model the second-order quantum coherence should remain unchanged. Note that at the single excitation level, one could anticipate situations in which the damping departs significantly from the classical model, for instance because of possible correlations between and within the different damping channels, such as excited phonons, background ion-cores, electron gas collisions and interband transition processes (involving electron-hole pairs). Indeed, closer to the plasmon wavelength, the SPP character becomes more electron-like and such effects may become important. This regime remains to be investigated.
\begin{figure}[thb]
\includegraphics[width=16cm]{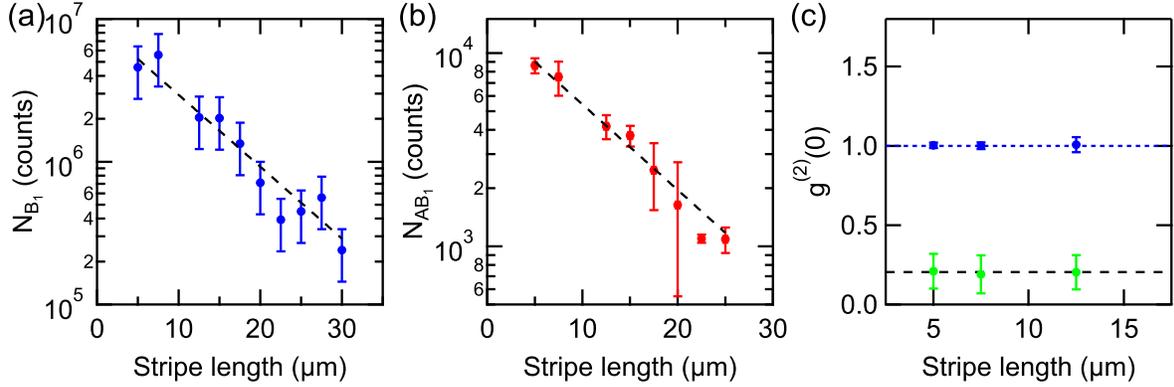}
\caption{Count rate statistics of the light out-coupled from the surface plasmon waveguides. (a) $N_{B_1}$ for injected attenuated laser as the length of the waveguide is increased. Dashed black line is a exponential fit, yielding a propagation length $\ell = \text{8.9}\pm\text{1.7~}\mu\text{m}$. (b) Same for $N_{AB_1}$ at zero delay for injected single photons. Here, $\ell = \text{9.8}\pm\text{0.6~}\mu\text{m}$. (c) $g^{(2)}_c(0)$ for single photons injected in waveguides of varying lengths (green) and unconditioned $g^{(2)}(0)$ for a laser injected in the waveguides (blue). The black dashed line indicates the value found for the single photon source, and the blue dotted line the classical limit.}
\label{figure3}
\end{figure}
To explore the quantum statistics in the presence of loss we first measured the mean excitation rate over a range of waveguide lengths. To do this we measured the counts $N_{B_1}$ at detector $B_1$ for an attenuated laser at a fixed intensity and then the conditional counts $N_{AB_1}$ at detectors $A$ and $B_1$ for the single-photon source. In both cases, the effect of loss from the beamsplitter in the analysis stage was included in the overall detection efficiency, enabling us to disregard the data from detector $B_2$. In Fig.~\ref{figure3}(a) we show~$N_{B_1}$ against waveguide length for the injected attenuated laser. Fig.~\ref{figure3}(b) presents $N_{AB_1}$ at zero delay for injected single photons and as the length of the waveguide is increased. The $N_{AB_1}$ trend matches $N_{B_1}$, providing evidence that the effect of loss on the field of the single SPPs is consistent with the classical exponential behavior. The SPP propagation length $\ell$, defined as the length at which the intensity (mean photon number $\langle n \rangle$) decreases to $1/e$ of its original value, extracted from Fig.~\ref{figure3}(a) is $\ell = \text{8.9}\pm\text{1.7~}\mu\text{m}$, a value similar to $\ell = \text{9.8}\pm\text{0.6~}\mu\text{m}$ obtained from Fig.~\ref{figure3}(b). Both values are in good agreement with each other, but smaller than the propagation length expected from Finite Element Method (FEM) calculations of 16.7$\mu$m, due to imperfections introduced by the fabrication of the waveguides.

In Fig.~\ref{figure3}(c), we show $g^{(2)}_c(0)$ for the out-coupled light for injected single photons as the length of the waveguide is increased. The value of $g^{(2)}_c(0)$ for the down-converted photons only is plotted as a dashed black line for reference. One can clearly see that indeed the values remain unchanged for the lengths investigated. These results therefore provide evidence for the validity of a linear uncorrelated Markovian loss model for SPP damping at the single quanta level. This complements well and goes beyond previous studies looking into the preservation of entanglement via localised plasmons~\cite{Altewischer2002,Moreno2004} and nonclassicality via long-range surface plasmons~\cite{Huck2009}, where elements of plasmon loss were considered.


\begin{figure}[t]
\includegraphics[width=14cm]{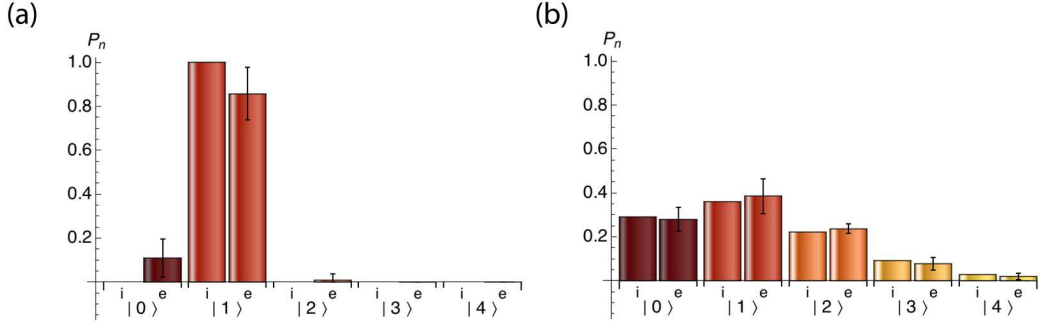}
\caption{Fock state populations of light out-coupled from the waveguides. (a) For conditioned single photons injected into the waveguides. (b) A laser attenuated to give on average one photon at the detection stage, $|\alpha|^2\sim1$. In both, the ideal populations are labeled $i$ and the experimentally reconstructed populations labeled $e$. For the attenuated laser, the ideal populations have been set to correspond to a weak coherent state with $|\alpha|^2=1.2$. In both plots the errors are calculated from a Monte-Carlo approach, propagating the errors from the measured data through the reconstruction algorithm.}
\label{figure4}
\end{figure}

We finish our investigation by probing the population structure of the conditioned SPP fields propagating along the waveguides. Note that this is the first time, to the best of our knowledge, that such a technique has been applied to a plasmonic quantum system. Here the number state, or Fock state populations $P_n$ for a given state, represents the probability of the state to have $n$ excitations. For a single-photon state $\ket{1}$ we have $P_1=1$ and $P_n=0,~\forall n \neq1$. On the other hand, for an attenuated laser described by the weak coherent state $\ket{\alpha}$, we have the Poissonian distribution $P_n=e^{-|\alpha|^2}\frac{|\alpha|^{2n}}{n!}$, where $|\alpha|^2=\langle n \rangle$ is the mean excitation number. For a mean excitation number of $|\alpha|^2=1$, we have the first six populations, $P_0=0.368$, $P_1=0.368$, $P_2=0.184$, $P_3=0.061$, $P_4=0.015$ and $P_5=0.003$, corresponding to the vacuum state $\ket{0}$ and number states $\ket{1}$, $\ket{2}$, $\ket{3}$, $\ket{4}$ and $\ket{5}$, respectively.

In Fig.~\ref{figure4}(a) we show a tomographic reconstruction of the Fock state populations of the field out-coupled from a $7.5\mu\text{m}$ long waveguide, for single photon excitation. In Fig.~\ref{figure4}(b) we show the reconstructed populations for a laser attenuated to give on average one photon at the detection stage, $|\alpha|^2\sim1$. Details of the tomographic reconstruction method~\cite{Zambra2005} are given in the supplementary information. In~Fig.~\ref{figure4}(b), one can clearly see the populations for the out-coupled light from the attenuated laser are consistent with a weak coherent state with mean excitation number of $\sim1$. On the other hand, the out-coupled light from the single-photon source, shown in~Fig.~\ref{figure4}(a), displays the strong presence of a single population, $P_1$, representative of a single excitation. This analysis complements the investigation performed in the previous two sections and confirms that we are exciting single SPPs on the waveguides when single photons are injected.


In this work we have used single photons generated by parametric down conversion to excite SPPs in metallic stripe waveguides. By measuring the second order quantum coherence function $g^{(2)}$ and the Fock state populations of the light coupled out of the gratings, we demonstrated the ability to excite single SPPs. Moreover, the effect of losses incurred during propagation of the single SPPs is consistent with the classical exponential behavior, and does not change the value of $g^{(2)}$, providing evidence that a linear uncorrelated Markovian loss model is valid for SPP damping at the single quanta level. Our results imply that building longer and more complex SPP waveguide structures operating in the quantum regime is realistic and opens up the possibility for future studies of new types of functioning devices based on quantum plasmonics.

{\acknowledgements S. A. M., M. S. K., M. T., S. K. C. and Y. S. acknowledge support by the UK Engineering and Physical Sciences Research Council (EPSRC) and the European Office for Aerospace Research and Development (EOARD). Y. S. and G.D. acknowledge funding from the Leverhulme Trust. S. K. O. thanks Prof. L. Yang and Prof. N. Imoto for their support.}



\begin{mcitethebibliography}{29}
\providecommand*\natexlab[1]{#1}
\providecommand*\mciteSetBstSublistMode[1]{}
\providecommand*\mciteSetBstMaxWidthForm[2]{}
\providecommand*\mciteBstWouldAddEndPuncttrue
  {\def\EndOfBibitem{\unskip.}}
\providecommand*\mciteBstWouldAddEndPunctfalse
  {\let\EndOfBibitem\relax}
\providecommand*\mciteSetBstMidEndSepPunct[3]{}
\providecommand*\mciteSetBstSublistLabelBeginEnd[3]{}
\providecommand*\EndOfBibitem{}
\mciteSetBstSublistMode{f}
\mciteSetBstMaxWidthForm{subitem}{(\alph{mcitesubitemcount})}
\mciteSetBstSublistLabelBeginEnd
  {\mcitemaxwidthsubitemform\space}
  {\relax}
  {\relax}

\bibitem[Takahara et~al.(1997)Takahara, Yamagishi, Taki, Morimoto, and
  Kobayashi]{Takahara1997}
Takahara,~J.; Yamagishi,~S.; Taki,~H.; Morimoto,~A.; Kobayashi,~T. \emph{Opt.
  Lett.} \textbf{1997}, \emph{22}, 475--477\relax
\mciteBstWouldAddEndPuncttrue
\mciteSetBstMidEndSepPunct{\mcitedefaultmidpunct}
{\mcitedefaultendpunct}{\mcitedefaultseppunct}\relax
\EndOfBibitem
\bibitem[Ebbesen et~al.(2008)Ebbesen, Genet, and Bozhevolnyi]{Ebbesen2008}
Ebbesen,~T.~W.; Genet,~C.; Bozhevolnyi,~S.~I. \emph{Phys. Today}
  \textbf{2008}, \emph{61}, 44 -- 50\relax
\mciteBstWouldAddEndPuncttrue
\mciteSetBstMidEndSepPunct{\mcitedefaultmidpunct}
{\mcitedefaultendpunct}{\mcitedefaultseppunct}\relax
\EndOfBibitem
\bibitem[Chang et~al.(2006)Chang, S\o{}rensen, Hemmer, and Lukin]{Chang2006}
Chang,~D.~E.; S\o{}rensen,~A.~S.; Hemmer,~P.~R.; Lukin,~M.~D. \emph{Phys. Rev.
  Lett.} \textbf{2006}, \emph{97}, 053002\relax
\mciteBstWouldAddEndPuncttrue
\mciteSetBstMidEndSepPunct{\mcitedefaultmidpunct}
{\mcitedefaultendpunct}{\mcitedefaultseppunct}\relax
\EndOfBibitem
\bibitem[Hong and Mandel(1986)Hong, and Mandel]{Hong1986}
Hong,~C.~K.; Mandel,~L. \emph{Phys. Rev. Lett.} \textbf{1986}, \emph{56},
  58--60\relax
\mciteBstWouldAddEndPuncttrue
\mciteSetBstMidEndSepPunct{\mcitedefaultmidpunct}
{\mcitedefaultendpunct}{\mcitedefaultseppunct}\relax
\EndOfBibitem
\bibitem[Akimov et~al.(2007)Akimov, Mukherjee, Yu, Chang, Zibrov, Hemmer, Park,
  and Lukin]{Akimov2007}
Akimov,~A.~V.; Mukherjee,~A.; Yu,~C.~L.; Chang,~D.~E.; Zibrov,~A.~S.;
  Hemmer,~P.~R.; Park,~H.; Lukin,~M.~D. \emph{Nature} \textbf{2007},
  \emph{450}, 402\relax
\mciteBstWouldAddEndPuncttrue
\mciteSetBstMidEndSepPunct{\mcitedefaultmidpunct}
{\mcitedefaultendpunct}{\mcitedefaultseppunct}\relax
\EndOfBibitem
\bibitem[Zuloaga et~al.(2009)Zuloaga, Prodan, and Nordlander]{Zuloaga2009}
Zuloaga,~J.; Prodan,~E.; Nordlander,~P. \emph{Nano Lett.} \textbf{2009},
  \emph{9}, 887--891\relax
\mciteBstWouldAddEndPuncttrue
\mciteSetBstMidEndSepPunct{\mcitedefaultmidpunct}
{\mcitedefaultendpunct}{\mcitedefaultseppunct}\relax
\EndOfBibitem
\bibitem[Artuso et~al.(2011)Artuso, Bryant, Garcia-Etxarri, and
  Aizpurua]{Artuso2011}
Artuso,~R.~D.; Bryant,~G.~W.; Garcia-Etxarri,~A.; Aizpurua,~J. \emph{Phys. Rev.
  B} \textbf{2011}, \emph{83}, 235406\relax
\mciteBstWouldAddEndPuncttrue
\mciteSetBstMidEndSepPunct{\mcitedefaultmidpunct}
{\mcitedefaultendpunct}{\mcitedefaultseppunct}\relax
\EndOfBibitem
\bibitem[Altewischer et~al.(2002)Altewischer, van Exter, and
  Woerdman]{Altewischer2002}
Altewischer,~E.; van Exter,~M.~P.; Woerdman,~J.~P. \emph{Nature} \textbf{2002},
  \emph{418}, 304\relax
\mciteBstWouldAddEndPuncttrue
\mciteSetBstMidEndSepPunct{\mcitedefaultmidpunct}
{\mcitedefaultendpunct}{\mcitedefaultseppunct}\relax
\EndOfBibitem
\bibitem[Moreno et~al.(2004)Moreno, Garc\'{\i}a-Vidal, Erni, Cirac, and
  Mart\'{\i}n-Moreno]{Moreno2004}
Moreno,~E.; Garc\'{\i}a-Vidal,~F.~J.; Erni,~D.; Cirac,~J.~I.;
  Mart\'{\i}n-Moreno,~L. \emph{Phys. Rev. Lett.} \textbf{2004}, \emph{92},
  236801\relax
\mciteBstWouldAddEndPuncttrue
\mciteSetBstMidEndSepPunct{\mcitedefaultmidpunct}
{\mcitedefaultendpunct}{\mcitedefaultseppunct}\relax
\EndOfBibitem
\bibitem[Fasel et~al.(2005)Fasel, Robin, Moreno, Erni, Gisin, and
  Zbinden]{Fasel2005}
Fasel,~S.; Robin,~F.; Moreno,~E.; Erni,~D.; Gisin,~N.; Zbinden,~H. \emph{Phys.
  Rev. Lett.} \textbf{2005}, \emph{94}, 110501\relax
\mciteBstWouldAddEndPuncttrue
\mciteSetBstMidEndSepPunct{\mcitedefaultmidpunct}
{\mcitedefaultendpunct}{\mcitedefaultseppunct}\relax
\EndOfBibitem
\bibitem[Fasel et~al.(2006)Fasel, Halder, Gisin, and Zbinden]{Fasel2006}
Fasel,~S.; Halder,~M.; Gisin,~N.; Zbinden,~H. \emph{New J. Phys.}
  \textbf{2006}, \emph{8}, 13\relax
\mciteBstWouldAddEndPuncttrue
\mciteSetBstMidEndSepPunct{\mcitedefaultmidpunct}
{\mcitedefaultendpunct}{\mcitedefaultseppunct}\relax
\EndOfBibitem
\bibitem[Huck et~al.(2009)Huck, Smolka, Lodahl, S\o{}rensen, Boltasseva,
  Janousek, and Andersen]{Huck2009}
Huck,~A.; Smolka,~S.; Lodahl,~P.; S\o{}rensen,~A.~S.; Boltasseva,~A.;
  Janousek,~J.; Andersen,~U.~L. \emph{Phys. Rev. Lett.} \textbf{2009},
  \emph{102}, 246802\relax
\mciteBstWouldAddEndPuncttrue
\mciteSetBstMidEndSepPunct{\mcitedefaultmidpunct}
{\mcitedefaultendpunct}{\mcitedefaultseppunct}\relax
\EndOfBibitem
\bibitem[Kolesov et~al.(2009)Kolesov, Grotz, Balasubramanian, Stöhr, Nicolet,
  Hemmer, Jelezko, and Wrachtrup]{Kolesov2009}
Kolesov,~R.; Grotz,~B.; Balasubramanian,~G.; Stöhr,~R.~J.; Nicolet,~A. A.~L.;
  Hemmer,~P.~R.; Jelezko,~F.; Wrachtrup,~J. \emph{Nature Phys.}
  \textbf{2009}, \emph{5}, 470\relax
\mciteBstWouldAddEndPuncttrue
\mciteSetBstMidEndSepPunct{\mcitedefaultmidpunct}
{\mcitedefaultendpunct}{\mcitedefaultseppunct}\relax
\EndOfBibitem
\bibitem[Heeres et~al.(2010)Heeres, Dorenbos, Koene, Solomon, Kouwenhoven, and
  Zwiller]{Heeres2010}
Heeres,~R.~W.; Dorenbos,~S.~N.; Koene,~B.; Solomon,~G.~S.; Kouwenhoven,~L.~P.;
  Zwiller,~V. \emph{Nano Lett.} \textbf{2010}, \emph{10}, 661--664\relax
\mciteBstWouldAddEndPuncttrue
\mciteSetBstMidEndSepPunct{\mcitedefaultmidpunct}
{\mcitedefaultendpunct}{\mcitedefaultseppunct}\relax
\EndOfBibitem
\bibitem[Chang et~al.(2007)Chang, Sorensen, Demler, and Lukin]{Chang2007}
Chang,~D.~E.; Sorensen,~A.~S.; Demler,~E.~A.; Lukin,~M.~D. \emph{Nature
  Phys.} \textbf{2007}, \emph{3}, 807\relax
\mciteBstWouldAddEndPuncttrue
\mciteSetBstMidEndSepPunct{\mcitedefaultmidpunct}
{\mcitedefaultendpunct}{\mcitedefaultseppunct}\relax
\EndOfBibitem
\bibitem[Burnham and Weinberg(1970)Burnham, and Weinberg]{Burnham1970}
Burnham,~D.~C.; Weinberg,~D.~L. \emph{Phys. Rev. Lett.} \textbf{1970},
  \emph{25}, 84--87\relax
\mciteBstWouldAddEndPuncttrue
\mciteSetBstMidEndSepPunct{\mcitedefaultmidpunct}
{\mcitedefaultendpunct}{\mcitedefaultseppunct}\relax
\EndOfBibitem
\bibitem[Berini(2001)]{Berini2001}
Berini,~P. \emph{Phys. Rev. B} \textbf{2001}, \emph{63}, 125417\relax
\mciteBstWouldAddEndPuncttrue
\mciteSetBstMidEndSepPunct{\mcitedefaultmidpunct}
{\mcitedefaultendpunct}{\mcitedefaultseppunct}\relax
\EndOfBibitem
\bibitem[Lamprecht et~al.(2001)Lamprecht, Krenn, Schider, Ditlbacher, Salerno,
  Felidj, Leitner, Aussenegg, and Weeber]{Lamprecht2001}
Lamprecht,~B.; Krenn,~J.~R.; Schider,~G.; Ditlbacher,~H.; Salerno,~M.;
  Felidj,~N.; Leitner,~A.; Aussenegg,~F.~R.; Weeber,~J.~C. \emph{App.
  Phys. Lett.} \textbf{2001}, \emph{79}, 51--53\relax
\mciteBstWouldAddEndPuncttrue
\mciteSetBstMidEndSepPunct{\mcitedefaultmidpunct}
{\mcitedefaultendpunct}{\mcitedefaultseppunct}\relax
\EndOfBibitem
\bibitem[Tame et~al.(2008)Tame, Lee, Lee, Ballester, Paternostro, Zayats, and
  Kim]{Tame2008}
Tame,~M.~S.; Lee,~C.; Lee,~J.; Ballester,~D.; Paternostro,~M.; Zayats,~A.~V.;
  Kim,~M.~S. \emph{Phys. Rev. Lett.} \textbf{2008}, \emph{101}, 190504\relax
\mciteBstWouldAddEndPuncttrue
\mciteSetBstMidEndSepPunct{\mcitedefaultmidpunct}
{\mcitedefaultendpunct}{\mcitedefaultseppunct}\relax
\EndOfBibitem
\bibitem[Zia et~al.(2005)Zia, Selker, and Brongersma]{Zia2005}
Zia,~R.; Selker,~M.~D.; Brongersma,~M.~L. \emph{Phys. Rev. B} \textbf{2005},
  \emph{71}, 165431\relax
\mciteBstWouldAddEndPuncttrue
\mciteSetBstMidEndSepPunct{\mcitedefaultmidpunct}
{\mcitedefaultendpunct}{\mcitedefaultseppunct}\relax
\EndOfBibitem
\bibitem[Brown and Twiss(1956)Brown, and Twiss]{Brown1956}
Brown,~R.~H.; Twiss,~R.~Q. \emph{Nature} \textbf{1956}, \emph{177}, 27\relax
\mciteBstWouldAddEndPuncttrue
\mciteSetBstMidEndSepPunct{\mcitedefaultmidpunct}
{\mcitedefaultendpunct}{\mcitedefaultseppunct}\relax
\EndOfBibitem
\bibitem[Loudon(2000)]{Loudon2000}
Loudon,~R. \emph{The Quantum Theory of Light, 3rd Edition}; Oxford University
  Press, 2000\relax
\mciteBstWouldAddEndPuncttrue
\mciteSetBstMidEndSepPunct{\mcitedefaultmidpunct}
{\mcitedefaultendpunct}{\mcitedefaultseppunct}\relax
\EndOfBibitem
\bibitem[Thorn et~al.(2004)Thorn, Neel, Donato, Bergreen, Davies, and
  Beck]{Thorn2004}
Thorn,~J.~J.; Neel,~M.~S.; Donato,~V.~W.; Bergreen,~G.~S.; Davies,~R.~E.;
  Beck,~M. \emph{Am. J. Phys.} \textbf{2004}, \emph{72},
  1210--1219\relax
\mciteBstWouldAddEndPuncttrue
\mciteSetBstMidEndSepPunct{\mcitedefaultmidpunct}
{\mcitedefaultendpunct}{\mcitedefaultseppunct}\relax
\EndOfBibitem
\bibitem[Razavi et~al.(2009)Razavi, Söllner, Bocquillon, Couteau, Laflamme, and
  Weihs]{Razavi2009}
Razavi,~M.; Söllner,~I.; Bocquillon,~E.; Couteau,~C.; Laflamme,~R.; Weihs,~G.
  \emph{J. Phys. B}
  \textbf{2009}, \emph{42}, 114013\relax
\mciteBstWouldAddEndPuncttrue
\mciteSetBstMidEndSepPunct{\mcitedefaultmidpunct}
{\mcitedefaultendpunct}{\mcitedefaultseppunct}\relax
\EndOfBibitem
\bibitem{Footnote1}
Here the response time of the detectors and time-tagging unit in our experiment, $\tau_d\sim350$~ps, is larger than the coherence time $\tau_c \approx 60$~fs of our single photon source~\cite{Bettelli2010,Hockel2011}
\EndOfBibitem
\bibitem[Bettelli(2010)]{Bettelli2010}
Bettelli,~S. \emph{Phys. Rev. A} \textbf{2010}, \emph{81}, 037801\relax
\mciteBstWouldAddEndPuncttrue
\mciteSetBstMidEndSepPunct{\mcitedefaultmidpunct}
{\mcitedefaultendpunct}{\mcitedefaultseppunct}\relax
\EndOfBibitem
\bibitem[H{\"o}ckel et~al.(2011)H{\"o}ckel, Koch, and Benson]{Hockel2011}
H{\"o}ckel,~D.; Koch,~L.; Benson,~O. \emph{Phys. Rev. A} \textbf{2011},
  \emph{83}, 013802\relax
\mciteBstWouldAddEndPuncttrue
\mciteSetBstMidEndSepPunct{\mcitedefaultmidpunct}
{\mcitedefaultendpunct}{\mcitedefaultseppunct}\relax
\EndOfBibitem
\bibitem[Blauensteiner et~al.(2009)Blauensteiner, Herbauts, Bettelli, Poppe,
  and H{\"u}bel]{Blauensteiner2009}
Blauensteiner,~B.; Herbauts,~I.; Bettelli,~S.; Poppe,~A.; H{\"u}bel,~H.
  \emph{Phys. Rev. A} \textbf{2009}, \emph{79}, 063846\relax
\mciteBstWouldAddEndPuncttrue
\mciteSetBstMidEndSepPunct{\mcitedefaultmidpunct}
{\mcitedefaultendpunct}{\mcitedefaultseppunct}\relax
\EndOfBibitem
\bibitem[Zayats et~al.(2005)Zayats, Smolyaninov, and Maradudin]{Zayats2005}
Zayats,~A.~V.; Smolyaninov,~I.~I.; Maradudin,~A.~A. \emph{Phys. Rep.}
  \textbf{2005}, \emph{408}, 131 -- 314\relax
\mciteBstWouldAddEndPuncttrue
\mciteSetBstMidEndSepPunct{\mcitedefaultmidpunct}
{\mcitedefaultendpunct}{\mcitedefaultseppunct}\relax
\EndOfBibitem
\bibitem[Zambra et~al.(2005)Zambra, Andreoni, Bondani, Gramegna, Genovese,
  Brida, Rossi, and Paris]{Zambra2005}
Zambra,~G.; Andreoni,~A.; Bondani,~M.; Gramegna,~M.; Genovese,~M.; Brida,~G.;
  Rossi,~A.; Paris,~M. G.~A. \emph{Phys. Rev. Lett.} \textbf{2005}, \emph{95},
  063602\relax
\mciteBstWouldAddEndPuncttrue
\mciteSetBstMidEndSepPunct{\mcitedefaultmidpunct}
{\mcitedefaultendpunct}{\mcitedefaultseppunct}\relax
\EndOfBibitem
\end{mcitethebibliography}

\begin{mcitethebibliography}{11}
\providecommand*\natexlab[1]{#1}
\providecommand*\mciteSetBstSublistMode[1]{}
\providecommand*\mciteSetBstMaxWidthForm[2]{}
\providecommand*\mciteBstWouldAddEndPuncttrue
  {\def\EndOfBibitem{\unskip.}}
\providecommand*\mciteBstWouldAddEndPunctfalse
  {\let\EndOfBibitem\relax}
\providecommand*\mciteSetBstMidEndSepPunct[3]{}
\providecommand*\mciteSetBstSublistLabelBeginEnd[3]{}
\providecommand*\EndOfBibitem{}
\mciteSetBstSublistMode{f}
\mciteSetBstMaxWidthForm{subitem}{(\alph{mcitesubitemcount})}
\mciteSetBstSublistLabelBeginEnd
  {\mcitemaxwidthsubitemform\space}
  {\relax}
  {\relax}

\bibitem[Loudon(2000)]{sLoudon2000}
Loudon,~R. \emph{The Quantum Theory of Light, 3rd Edition}; Oxford University
  Press, 2000\relax
\mciteBstWouldAddEndPuncttrue
\mciteSetBstMidEndSepPunct{\mcitedefaultmidpunct}
{\mcitedefaultendpunct}{\mcitedefaultseppunct}\relax
\EndOfBibitem
\bibitem[Hong and Mandel(1986)Hong, and Mandel]{sHong1986}
Hong,~C.~K.; Mandel,~L. \emph{Phys. Rev. Lett.} \textbf{1986}, \emph{56},
  58--60\relax
\mciteBstWouldAddEndPuncttrue
\mciteSetBstMidEndSepPunct{\mcitedefaultmidpunct}
{\mcitedefaultendpunct}{\mcitedefaultseppunct}\relax
\EndOfBibitem
\bibitem[Burnham and Weinberg(1970)Burnham, and Weinberg]{sBurnham1970}
Burnham,~D.~C.; Weinberg,~D.~L. \emph{Phys. Rev. Lett.} \textbf{1970},
  \emph{25}, 84--87\relax
\mciteBstWouldAddEndPuncttrue
\mciteSetBstMidEndSepPunct{\mcitedefaultmidpunct}
{\mcitedefaultendpunct}{\mcitedefaultseppunct}\relax
\EndOfBibitem
\bibitem[Razavi et~al.(2009)Razavi, Söllner, Bocquillon, Couteau, Laflamme, and
  Weihs]{sRazavi2009}
Razavi,~M.; Söllner,~I.; Bocquillon,~E.; Couteau,~C.; Laflamme,~R.; Weihs,~G.
  \emph{J. Phys. B}
  \textbf{2009}, \emph{42}, 114013\relax
\mciteBstWouldAddEndPuncttrue
\mciteSetBstMidEndSepPunct{\mcitedefaultmidpunct}
{\mcitedefaultendpunct}{\mcitedefaultseppunct}\relax
\EndOfBibitem
\bibitem[Baek and Kim(2008)Baek, and Kim]{sBaek2008}
Baek,~S.-Y.; Kim,~Y.-H. \emph{Phys. Rev. A} \textbf{2008}, \emph{77},
  043807\relax
\mciteBstWouldAddEndPuncttrue
\mciteSetBstMidEndSepPunct{\mcitedefaultmidpunct}
{\mcitedefaultendpunct}{\mcitedefaultseppunct}\relax
\EndOfBibitem
\bibitem[Thorn et~al.(2004)Thorn, Neel, Donato, Bergreen, Davies, and
  Beck]{sThorn2004}
Thorn,~J.~J.; Neel,~M.~S.; Donato,~V.~W.; Bergreen,~G.~S.; Davies,~R.~E.;
  Beck,~M. \emph{Am. J. Phys.} \textbf{2004}, \emph{72},
  1210--1219\relax
\mciteBstWouldAddEndPuncttrue
\mciteSetBstMidEndSepPunct{\mcitedefaultmidpunct}
{\mcitedefaultendpunct}{\mcitedefaultseppunct}\relax
\EndOfBibitem
\bibitem[Zambra et~al.(2005)Zambra, Andreoni, Bondani, Gramegna, Genovese,
  Brida, Rossi, and Paris]{sZambra2005}
Zambra,~G.; Andreoni,~A.; Bondani,~M.; Gramegna,~M.; Genovese,~M.; Brida,~G.;
  Rossi,~A.; Paris,~M. G.~A. \emph{Phys. Rev. Lett.} \textbf{2005}, \emph{95},
  063602\relax
\mciteBstWouldAddEndPuncttrue
\mciteSetBstMidEndSepPunct{\mcitedefaultmidpunct}
{\mcitedefaultendpunct}{\mcitedefaultseppunct}\relax
\EndOfBibitem
\bibitem[Simon et~al.(1974)Simon, Mitchell, and Watson]{sSimon1974}
Simon,~H.~J.; Mitchell,~D.~E.; Watson,~J.~G. \emph{Phys. Rev. Lett.}
  \textbf{1974}, \emph{33}, 1531--1534\relax
\mciteBstWouldAddEndPuncttrue
\mciteSetBstMidEndSepPunct{\mcitedefaultmidpunct}
{\mcitedefaultendpunct}{\mcitedefaultseppunct}\relax
\EndOfBibitem
\bibitem[Tame et~al.(2008)Tame, Lee, Lee, Ballester, Paternostro, Zayats, and
  Kim]{sTame2008}
Tame,~M.~S.; Lee,~C.; Lee,~J.; Ballester,~D.; Paternostro,~M.; Zayats,~A.~V.;
  Kim,~M.~S. \emph{Phys. Rev. Lett.} \textbf{2008}, \emph{101}, 190504\relax
\mciteBstWouldAddEndPuncttrue
\mciteSetBstMidEndSepPunct{\mcitedefaultmidpunct}
{\mcitedefaultendpunct}{\mcitedefaultseppunct}\relax
\EndOfBibitem
\bibitem[Elson and Ritchie(1971)Elson, and Ritchie]{sElson1971}
Elson,~J.~M.; Ritchie,~R.~H. \emph{Phys. Rev. B} \textbf{1971}, \emph{4},
  4129--4138\relax
\mciteBstWouldAddEndPuncttrue
\mciteSetBstMidEndSepPunct{\mcitedefaultmidpunct}
{\mcitedefaultendpunct}{\mcitedefaultseppunct}\relax
\EndOfBibitem
\bibitem[Zayats et~al.(2005)Zayats, Smolyaninov, and Maradudin]{sZayats2005}
Zayats,~A.~V.; Smolyaninov,~I.~I.; Maradudin,~A.~A. \emph{Phys. Rep.}
  \textbf{2005}, \emph{408}, 131 -- 314\relax
\mciteBstWouldAddEndPuncttrue
\mciteSetBstMidEndSepPunct{\mcitedefaultmidpunct}
{\mcitedefaultendpunct}{\mcitedefaultseppunct}\relax
\EndOfBibitem
\bibitem[Zia et~al.(2005)Zia, Selker, and Brongersma]{sZia2005}
Zia,~R.; Selker,~M.~D.; Brongersma,~M.~L. \emph{Phys. Rev. B} \textbf{2005},
  \emph{71}, 165431\relax
\mciteBstWouldAddEndPuncttrue
\mciteSetBstMidEndSepPunct{\mcitedefaultmidpunct}
{\mcitedefaultendpunct}{\mcitedefaultseppunct}\relax
\EndOfBibitem
\end{mcitethebibliography}


\providecommand*\mcitethebibliography{\thebibliography}
\csname @ifundefined\endcsname{endmcitethebibliography}
  {\let\endmcitethebibliography\endthebibliography}{}

\newpage
\setcounter{equation}{0}
\setcounter{figure}{0}

\section{Supplementary Information}

\subsection{1. Second-order quantum coherence function $g^{(2)}(\tau)$}

For quantized electromagnetic fields propagating in the $x$-direction with an arbitrary lateral beam profile, and represented by the electric field operator $\hat{E}^+(x,t)$, we have at a fixed position, $x=0$, the following definition~\cite{sLoudon2000}

\begin{equation}
g^{(2)}(\tau)=\frac{\langle \hat{E}^-(0)\hat{E}^-(\tau)\hat{E}^+(\tau)\hat{E}^+(0)\rangle}{\langle \hat{E}^-(0)\hat{E}^+(0)\rangle^2}.
\label{g2}
\end{equation}
Here $\langle\hat{X}\rangle$ represents the expectation value of the operator $\hat{X}$ with respect to the initial state of the field, {\it i.e.} an averaging over ensembles. The average of the intensity of the field is assumed to be constant over time, $\langle \hat{E}^-(\tau)\hat{E}^+(\tau)\rangle=\langle \hat{E}^-(0)\hat{E}^+(0)\rangle$.
Throughout we will suppress the position dependence of $\hat{E}^+(x,t)$, as $x$ is fixed at zero.
At zero time delay, $\tau=0$, for $n$-excitation states $\ket{n}$, we have that $\bra{n} \hat{E}^-(0)\hat{E}^-(0)\hat{E}^+(0)\hat{E}^+(0)\ket{n}=n(n-1)$ and $\bra{n} \hat{E}^-(0)\hat{E}^+(0)\ket{n}=n$, leading to the relation $g^{(2)}(0)=1-1/n$. In particular, for $n=1$ (single excitations), we have $g^{(2)}(0)=0$. Similarly, for $n=2$, $g^{(2)}(0)=0.5$: a measured value of $g^{(2)}(0)$ between 0 and 0.5 is a confirmation that we are dealing with single excitations. On the other hand, for attenuated laser light described by a weak coherent state $\ket{\alpha}=\sum_{n=0}^{\infty}e^{-|\alpha|^2}\frac{|\alpha|^{2n}}{n!}\ket{n}$, where $|\alpha|^2=\langle n \rangle$ is the mean excitation number, we have $g^{(2)}(0)=1$. Moreover, it can be shown using the Cauchy-Schwartz inequality that for any classical electromagnetic field, due to the absence of operators and their commutation relations for the classical electric field $E^+(x,t)$, the numerator in Eq.~\ref{g2} factorizes to give the inequality $g^{(2)}(0)\geq1$. Thus by measuring $g^{(2)}(0)$ for a given field, we can determine whether or not it is in the nonclassical regime ($g^{(2)}(0)<1$).

\subsection{2. Spontaneous parametric down-conversion and heralded single photons}
	\subsubsection{a) Heralded single photons}
	
The interaction Hamiltonian for type-I spontaneous parametric down-conversion is given by~\cite{sHong1986,sBurnham1970}

\begin{equation}
\hat{H}_I = \hbar ~\xi~ \hat{a}^\dag_A\hat{a}^\dag_B + H.c.
\end{equation}
Here, $\xi \propto \chi^{(2)}{\cal E}_p$,  where $\chi^{(2)}$ is the second-order nonlinear susceptibility of the BBO crystal in our setup and ${\cal E}_p$ is the amplitude of the classical coherent laser pump field. In addition, $\hat{a}^\dag_A$ ($\hat{a}^\dag_B$) is a creation operator for a photon in mode $A$ ($B$) and $H.c.$ represents the Hermitian conjugate. Taking the initial state of modes $A$ and $B$ to be the vacuum $\ket{\psi(0)}=\ket{0}_{AB}=\ket{0}_{A}\ket{0}_{B}$ and evolving it according to the Schr\"odinger equation as 

\begin{equation}
\ket{\psi(t)}=e^{-it\hat{H}_I/\hbar}\ket{\psi(0)},
\end{equation}
we obtain, up to first order in time, the state
\begin{equation}
\ket{\psi(t)}=(1-\mu^2/2)\ket{0}_{A}\ket{0}_{B}-i\mu\ket{1}_{A}\ket{1}_{B},
\end{equation}
where $\mu=\xi t$. By detecting a photon in mode $A$ we remove the first (vacuum) term and `herald' the presence of a single-photon state $\ket{1}_B$ in mode $B$, up to first order. By tuning the pump laser intensity appropriately, higher order terms can be made negligible in mode $B$, even if the detection in mode $A$ is not photon number resolving. Thus, we can use type-I SPDC to produce high-quality single photon states, $\ket{1}$, with larger generation rates than currently achieved with emitter-type sources, such as quantum dots~\cite{sRazavi2009}. 
\subsubsection{b) Characterization of our single photon source}	
\begin{figure}[h]
		\includegraphics[width=16cm]{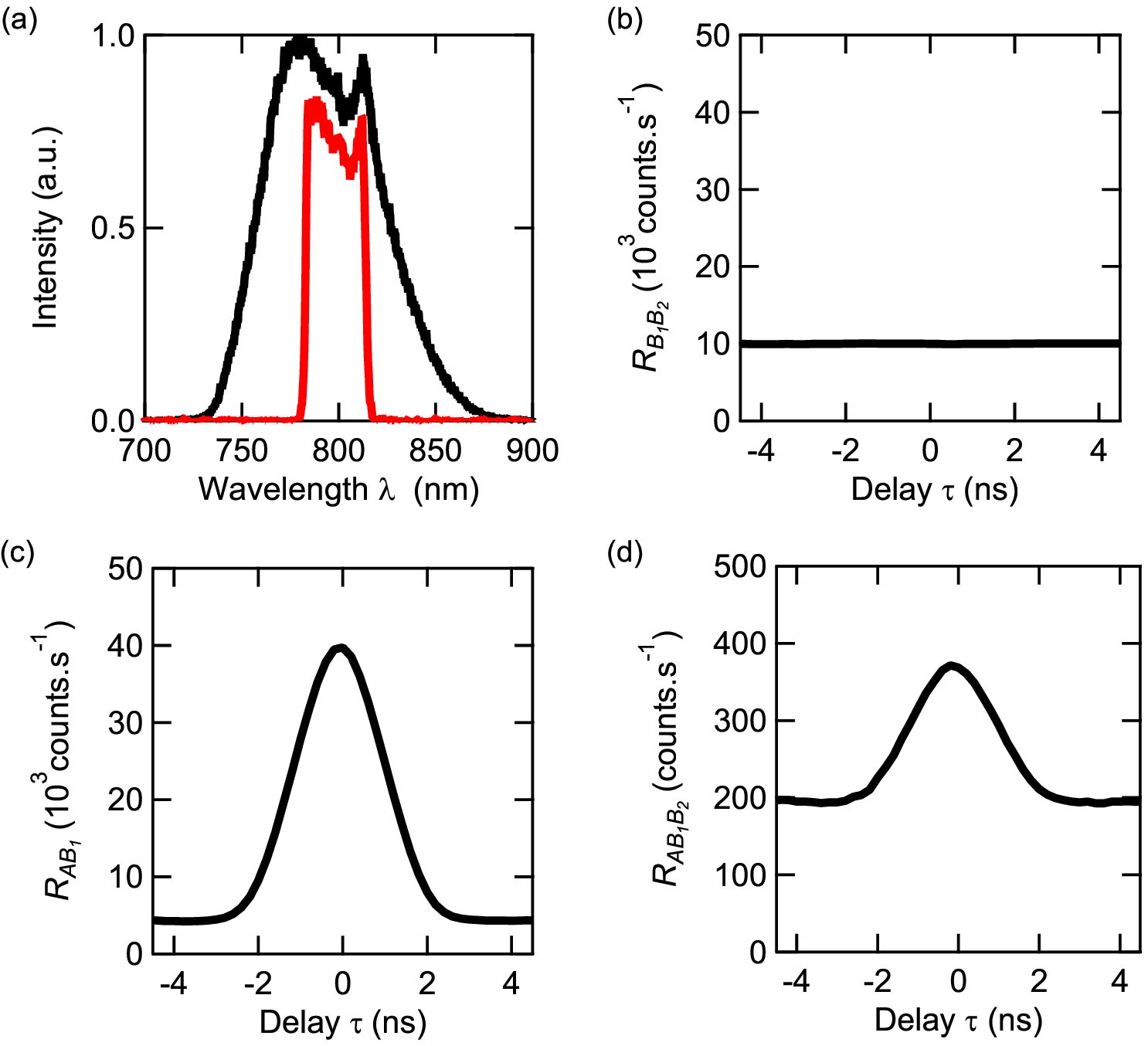}
			\caption{Characterization of heralded single-photon source. (a) Unconditioned spectral intensity function for photons in arms $A$ (red) and $B$ (black). Note that a $\lambda = 800$~nm bandpass filter, with bandwidth $\Delta\lambda = 30$~nm was placed in arm $A$ to spectrally select only a fraction of the down-converted photons. This effectively selects out an equivalent bandwidth in arm $B$ when the detections are conditioned on $A$. (b) (resp (c)) Coincidence rate between $B_1$ and $B_2$ (resp $A$ and $B_1$). (d) Triple coincidence rate $R_{AB_1B_2}$.}
		\label{figure5}
	\end{figure}

In the down conversion process, the phase matching conditions are not perfect in the experiment. For this reason, the down-converted light is not monochromatic, but presents a spectrum with finite width~\cite{sBaek2008}: this effect is observed with our source as shown in Fig.~\ref{figure5}(a), where we present its spectral properties. In Fig.~\ref{figure5}(b) the unconditioned coincidence rates $R_{B_1B_2}=N_{B_1B_2}/T$ (where $T$ is the integration time) are shown. No correlation between those two arms arises. On the other hand, when we plot the conditioned rates $R_{AB_1}=N_{AB_1}/T$ (Fig.~\ref{figure5}(c)) it is apparent that there is a very strong correlation between arm $B_1$ and the reference arm $A$ at zero delay. The same occurs for $B_2$. This figure shows as well that the configuration used allows for a single photon generation rate of about 10$^6$~s$^{-1}$.\\
\indent In addition, the triple coincidence rate $R_{AB_1B_2}=N_{AB_1B_2}/T$ is shown in Fig.~\ref{figure5}(d). On first thought, one would expect to see a value of zero at zero delay, as a coincidence between $A$ and $B_1$ should indicate the presence of a single photon in the system, and thus forbidding any simultaneous detection on $B_2$. However, there is a peak in $R_{AB_1B_2}$ at zero delay: this is due to the fact that we count coincidences within a finite time window $\Delta t$. In our case, $\Delta t=2\text{~ns}$: two "clicks" detected within a 2~ns-wide time window are considered as a coincidence, even if they are not exactly simultaneous. For this reason, accidental coincidences are measured, at a rate determined solely by the count rates on each detector, the integration time $T$ and the time window $\Delta t$. One can show that~\cite{sThorn2004}, if $R_{B_1}$ and $R_{B_2}$ are the single count rates at $B_1$ and $B_2$ respectively, the accidental coincidence rate at zero delay for three detectors $R_{acc}(0)$ is:
\begin{equation}
R_{acc}(0)=\Delta t R_{AB_1}R_{B_2}+\Delta t R_{AB_2}R_{B_1}.
\label{AccRates3d}
\end{equation}
The rate of triples observed at zero delay agree very well with the value expected for solely accidental coincidences. Additionally, these accidental coincidences lead to a value of $g^{(2)}(0)$ higher than zero. One can show~\cite{sThorn2004} that the offset on $g^{(2)}(0)$ due to accidental coincidences is:
\begin{equation}
g^{(2)}_{acc}(0)=\Delta t R_A \left(\frac{R_{B_1}}{R_{AB_1}}+\frac{R_{B_2}}{R_{AB_2}}\right).
\label{g2accidentals}
\end{equation}
By using the values observed in Fig.~\ref{figure5} and the count rates at each detector, one finds the value of $g^{(2)}(0)=0.23$ for our single photon source for $\Delta t=2\text{~ns}$, which is the value observed in Fig.~2(b) in the main text: the non-zero value of $g^{(2)}(0)$ is solely due to accidental coincidences.

\newpage
\newpage

\subsection{3. Fock state population tomography}
\begin{figure}[h]
\includegraphics[width=16cm]{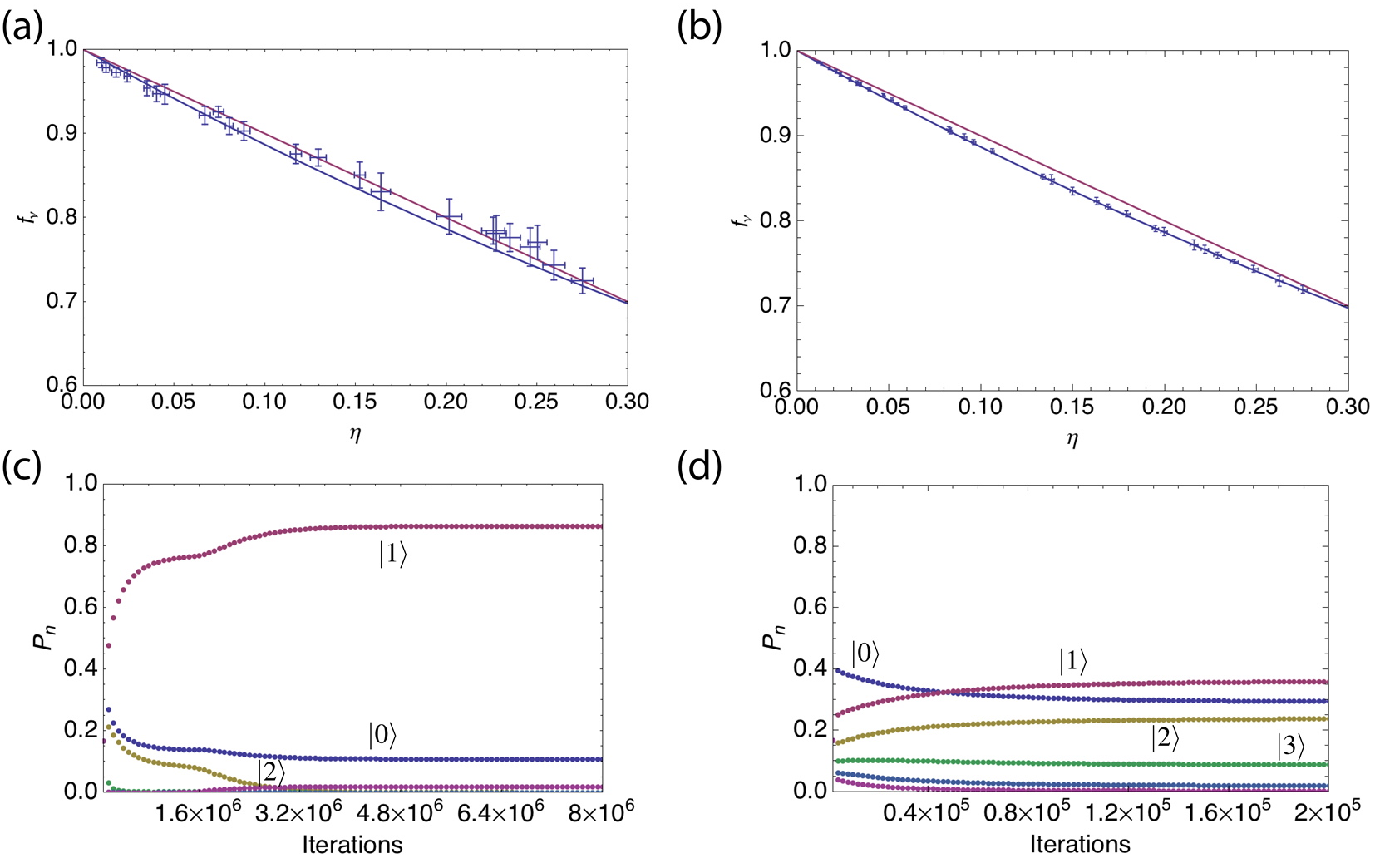}
\caption{No-click frequencies $f_\nu$ for reconstructing the Fock state populations. (a) No-click frequencies as detector efficiency $\eta_\nu$ is increased for single-photon data. (b) No-click frequencies as $\eta_\nu$ is increased for the attenuated laser. In both, the red line corresponds to an ideal single-photon and the blue curve corresponds to a weak coherent state with mean photon number $|\alpha|^2=1.2$. (c) and (d) show how the individual populations $P_n=\rho_n$ converge as the number of iterations of the algorithm increases for single photons and attenuated laser respectively.}
\label{figure7}
\end{figure}

Here we provide details of the tomographic method used to reconstruct the populations. We used the technique of Zambra {\it et al.}~\cite{sZambra2005} to measure the photon statistics based on on/off detection. An arbitrary quantum state can be written in the number basis as $\rho=\sum_{nm} \rho_{nm }\ket{n}\bra{m}$, where the diagonal elements $\rho_{nn}=\rho_{n}=\bra{n}\rho\ket{n}$ give the photon number distribution of the state $\rho$. Let $\eta$ be the efficiency of a given detector, so that $\eta$ is the probability for a single photon to be revealed and $(1-\eta)$ is the probability for it not to be revealed. Thus, the total probability of the detector not giving a `click' is $p(\eta)=\sum_n(1-\eta)^n\rho_n$.
Consider now a set of such detectors with different efficiencies, $\eta_\nu$. We then have $p(\eta_\nu)=\sum_n(1-\eta_\nu)^n\rho_n$, or more compactly written 
\begin{equation}
p_\nu=\sum_nA_{\nu n} \rho_n. 
\label{linpos}
\end{equation}
Here, $p_\nu$ can be obtained experimentally and $A_{\nu n}$ can be set by artificially changing the efficiency of the detector, leaving $\rho_n$ as the unknown parameter. Then, by assuming the $\rho_n$'s are negligible for $n>n_t$, where $n_t$ is a truncation number and we have at least $N>n_t$ detector efficiencies, Eq.~\ref{linpos} is a LINPOS problem~\cite{sZambra2005} and one can use the expectation maximization (EM) algorithm, which converges to the maximum likelihood solution. By imposing the physical constraint $\sum_n \rho_n=1$ we have the iterative solution
\begin{equation}
\rho_n^{(i+1)}=\rho_n^{(i)}\sum_{\nu}^N\frac{A_{\nu n}}{\sum_\lambda A_{\lambda n}}\frac{f_\nu}{p_\nu[\{ \rho_n^{(i)}\}]},
\end{equation}
where $\rho_n^{(i)}$ is the value of $\rho_n$ evaluated at the $i$-th iteration, $f_\nu$ are the experimental frequencies of the `no-click' events for $\eta=\eta_\nu$ (whose ideal values are $p_\nu$) and $p_\nu[\{ \rho_n^{(i)}\}]$ are the probabilities $p_\nu$ calculated using the reconstructed distribution $\{ \rho_n^{(i)}\}$ at the $i$-th iteration, {\it i.e.} $p_\nu[\{ \rho_n^{(i)}\}]=\sum_n^{n_t}(1-\eta_\nu)^n\rho_n^{(i)}$. We start the iterations with the unbiased distribution $\rho_n^{(0)}=1/(1+n_t)$ and use the experimental results $f_\nu=n_{0,\nu}/n_\nu$, where for a given detector efficiency $\eta_\nu$, $n_\nu$ is the total number of runs (state preparation and measurement) and $n_{0,\nu}$ is the number of no-click events for these runs. The EM algorithm is then carried out until the changes in the population numbers $\rho_n$ between iterations reduce below a given threshold, $\epsilon$. For the field out-coupled from the waveguides for the single photon source, we have $n_\nu=N_A$ and $n_{0,\nu}=N_A-N_{AB_{1,\nu}}\eta_x$. Here, $\eta_x$ is a loss scaling factor given by $\eta_x=\eta_d/\eta_0$, which allows us to consider the tomography being performed on the state that enters the detection and analysis stage in our setup (rather than a tomography of the initial state generated, as carried out by Zambra {\it et al.}~\cite{sZambra2005}). The loss $\eta_d=0.55/2$ corresponds to detector $B_1$'s intrinsic efficiency around the operating wavelength of the field used ($\lambda=$808nm), combined with that of the beamsplitter in front of it. The loss $\eta_0=N_{AB_{1,0}}/N_A$ is the total loss from initial state generation to detection at $B_1$. To measure the coincidences $N_{AB_{1,\nu}}$ we have set the coincidence window to $\Delta t=2$ns. A set of efficiencies are then introduced using an ND filter wheel. Here, the efficiencies $\eta_\nu=\eta_dN_{B_1,\nu}/N_{B_{1,0}}$. For the attenuated laser source, as it is not based on conditional measurements at detector $A$, we set a window of 500~ns for a measurement duration every 10~$\mu$s and carry out 10,000 runs. Thus $n_\nu=10,000$ and $n_{0,\nu}=10,000-N_{AB_{1,\nu}}$, where $N_{AB_{1,\nu}}$ is the total number of clicks from the 10,000 runs. The efficiencies $\eta_\nu=\eta_dN_{B_1,\nu}/N_{B_{1,0}}$, where $\eta_d=0.55/2$ is used as before.
In Fig.~\ref{figure7} we show the dependence of the no-click frequencies $f_\nu$ with the detector efficiencies $\eta_\nu$ measured in our experiment for the light out-coupled from the waveguides for the single-photon source and that for the attenuated laser. In both plots, background counts were subtracted from the singles ($N_A$ and $N_{B_1}$) and doubles ($N_{AB}$ and $N_{AB_1}$) at the detectors. Using these plots, the reconstructed populations from the EM algorithm are found and shown in the main text in Fig.~4. To calculate the errors in the populations we used a Monte-Carlo approach, propagating the errors from the measured data shown in Fig.~\ref{figure7} through the reconstruction algorithm, with Gaussian distributions placed on the values.

\newpage
\newpage

\subsection{4. Grating coupling}

Here we provide a theoretical model to describe the transfer of single photons to single SPPs via the grating coupling method used in our experiment. At the single-photon level only small intensities of the photon field are involved and therefore any nonlinear terms in the photon-SPP coupling can be effectively neglected~\cite{sSimon1974}, leading to the following linear coupling Hamiltonian~\cite{sTame2008}
\begin{eqnarray}
\hat{H}&=& \int_{0}^{\infty}{\mathrm d} \omega \hbar \omega \hat{a}^{\dag}(\omega)\hat{a}(\omega)+\int_{0}^{\infty}{\mathrm d} \omega \hbar \omega \hat{b}^{\dag}(\omega)\hat{b}(\omega)  \label{Hamil} \\
& & + i \hbar \int_{0}^{\infty}{\mathrm d} \omega [g(\omega)\hat{a}^{\dag}(\omega)\hat{b}(\omega)-g^*(\omega)\hat{b}^{\dag}(\omega)\hat{a}(\omega)]. \nonumber
\end{eqnarray}
Here, the $\hat{a}(\omega)$'s ($\hat{a}^{\dag}(\omega)$'s) correspond to annihilation (creation) operators for the photons which obey bosonic commutation relations $[\hat{a}(\omega),\hat{a}^{\dag}(\omega')]=\delta(\omega-\omega')$. Due to the collective nature of the electron charge density waves and the frequency regime of our experiment, a macroscopic approach for the resulting electromagnetic field is appropriate for the SPPs~\cite{sElson1971}. Upon quantization, they are therefore assumed to behave as bosonic modes. Thus, the $\hat{b}(\omega)$'s ($\hat{b}^{\dag}(\omega)$'s) correspond to annihilation (creation) operators for the SPPs which should, in principle, obey bosonic commutation relations $[\hat{b}(\omega),\hat{b}^{\dag}(\omega')]=\delta(\omega-\omega')$. In the Hamiltonian given in Eq.(9), the first and second terms are the photon and SPP fields' free-energy respectively. 
The last term describes interactions between the two fields, where the coupling $g(\omega)$ is proportional to the overlap of the field scattered by the grating with the SPP mode on the waveguide stripe. Carrying out the overlap integrals for the fields in a particular scenario gives the phase-matching conditions for the coupling and the coupling value itself. For perfect coupling, {\it i.e.} a single photon injected into the grating scatters and couples to a single SPP with unit efficiency, we have $g(\omega)=\pi/2$~\cite{sTame2008}. Thus, the geometry of the grating should be optimised to achieve a coupling as close as possible to $\pi/2$, with negligible deviation over the bandwidth, $\Delta \omega$, of the incoming photon - in order to avoid significant wavepacket distortion and loss during the transfer process. A first approximation for the phase-matching condition for photons injected normal to the surface and with TM polarization (with respect to the surface plane) is given by~\cite{sZayats2005} $k_{sp}=k_g$, where $k_{sp}$ is the magnitude of the SPP wavevector in the direction of propagation along the waveguide and $k_g=\frac{2 \pi}{\Lambda}m$ is the grating momentum ($\Lambda$ is the grating period and $m$ is an integer). For the metallic strip in our experiment, we use the approximation~\cite{sZia2005}, $k_{sp}=\sqrt{\frac{\omega^2}{c^2}\frac{\epsilon_m}{1+\epsilon_m}-\frac{\pi^2}{W^2}}$, where $\epsilon_m$ is the permittivity of gold, $W=3\mu$m is the width of the waveguide and $\omega$ corresponds to a free space wavelength of $\lambda_0=808$~nm. Taking the $m=+1$ grating momentum, one finds a grating period of $\Lambda=802$nm. We use this as our starting point and perform FEM simulations to optimise the coupling by modifying the height, width, period and number of grooves for the grating on the waveguide. We find the optimal period of $\Lambda=680$~nm. Such a large deviation from the approximate result can be explained by the use of deep grooves in our gratings, as the phase-matching condition $k_{sp}=k_g$ is only approximate for shallow gratings~\cite{sZayats2005} (weak perturbations). 

\providecommand*\mcitethebibliography{\thebibliography}
\csname @ifundefined\endcsname{endmcitethebibliography}
  {\let\endmcitethebibliography\endthebibliography}{}

\end{document}